# Lets Talk About Black Hole Singularities

Does the collision of black hole singularities imprint an observable quantum signature on the resulting gravitational wave signal?

______

By Abraham Loeb on May 14, 2018

The singularities at the centers of astrophysical black holes mark the breakdown of Einstein's theory of gravity, General Relativity. They represent the only breakdown sites accessible to experimentalists, since the other known singularity, the Big Bang, is believed to be invisible due to the vast expansion that occurred afterwards during cosmic inflation.

Every physicist knows these facts but very few discuss black hole singularities, as if the topic is taboo. The reason is simple. To explore the true nature of singularities we need a theory that unifies General Relativity with Quantum Mechanics and we do not have a unique, well defined formalism for doing that. Even in the context of specific proposals for a unified model, such as String Theory, the nature of black hole singularities is rarely discussed because of its mathematical complexity. But perhaps the time is ripe now to open up this discussion, given that the 2017 Nobel Prize was awarded to the LIGO team for discovering gravitational waves from collisions of black holes[1]. An observable quantum signal from the embedded singularities could guide us in the search for a unified theory. This thought occurred to me during two back-to-back conferences that we hosted at Harvard University on May 7-11, 2018, one[2] on "Gravitational Wave Astrophysics" and the second[3] being the annual conference of Harvard's *Black Hole Initiative*. A few days earlier, the basement at my home was flooded since the sewer pipe was clogged by tree roots, and the five hours spent with a plumber in fixing this problem led me to realize that any water going down the drain collects somewhere. Usually the sewer pipe takes the water to a town reservoir and we do not think about where it goes because we do not see the water once it leaves our property. But because the sewer pipe at my home was clogged, the water flooded my basement and then I started thinking about the analogous problem of where the matter that makes a black hole collects. The reservoir in that case is the singularity.

True, the singularity of a stationary black hole is hidden behind an event horizon for any external observer. This "Cosmic Censorship"[4] is a good reason for ignoring the observational consequences of singularities when probing the calm spacetime around isolated black holes, for example - while imaging the silhouette of Sagittarius A* at the center of the Milky Way with the Event Horizon Telescope. But this does not imply that observers, more generally, can never study empirically the nature of the singularities. When children get a birthday present wrapped in a box , they attempt to learn about its nature without seeing it directly by shaking the box and listening to its vibrations. Similarly, we can listen to the vibrations of a black hole horizon that is strongly shaken through its collision with another black hole, hoping to learn more about the nature of the singularities hidden inside. Future generations of LIGO detectors could serve as the "child's ears" in extracting new information from these vibrations.

A particularly interesting question is what happens when two singularities collide. How do they merge to a single singularity and does this process have an imprint on the gravitational wave signal that is observable by LIGO? Naively, one might argue that computer simulations have already calculated the gravitational wave signals from black hole collisions and there is no hint for the content of the "event horizon box" in these signals. Existing simulations cut out completely the region around the singularities by postulating that this region will not have observable effects (and justifiably so within General Relativity alone), but they do not incorporate quantum mechanical modifications of General Relativity which could link the fate of the excised region with the rest of spacetime. If there are observable signatures of merging singularities, existing computer simulations are, by construction, blind to them[5].

What would a singularity look like in the quantum mechanical context? Most likely, it would appear as an extreme concentration of a huge mass (more than a few solar masses for astrophysical black holes) within a tiny volume. The size of the reservoir that drains all matter which fell into an astrophysical black hole is unknown. We could envision the remnant singularity as being a finite-size reservoir in equilibrium, similar to the halos of galaxies, where the motion of infalling particles is turned around and confined by their binding gravitational attraction. One might assume that the outer boundary of the remnant object is some small fraction of its Schwarzschild radius, $2GM/c^2$ (which equals 3 kilometers times the black hole mass, M, in solar mass units), corresponding to a universal curvature scale at which Einstein's theory of gravity breaks down due to quantum corrections. In case the density of the object is universal (set, for example, by the Planck energy density), the size of the object that replaces the singularity will scale with its mass over the Planck mass ($10^{-5}$ grams) to the power of 1/3 times the Planck length ($10^{-33}$ centimeter, or $10^{-20}$ times the size of the proton).

Now imagine two such singularities colliding as a result of the merger of two black holes. Although the collision of these objects might not be visible directly to an external observer (unless a "naked singularity" appears during the process), the interesting question is whether the collision will produce a transient burst of energy that is observable to the outside world through the vibrations it induces in the event horizon. Is that possible?

This is an extremely interesting question that should be discussed further. It could motivate gravitational wave observers to develop more sensitive detectors. At the very least, we might be able to outline the landscape of possibilities. Science is work in progress and most of the fun exploring it is spent in uncharted territories.

I often encourage my string-theory colleagues to contemplate testing their theory by boarding a futuristic spacecraft that will take them into the event horizon of a nearby black hole. Perhaps future extensions of LIGO can save them the expense of this lengthy one-way trip.

## ABOUT THE AUTHOR

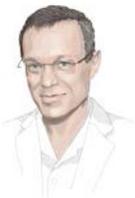

**Abraham Loeb**

Abraham Loeb is chair of the astronomy department at Harvard University, founding director of Harvard's Black Hole Initiative and director of the Institute for Theory and Computation at the Harvard-Smithsonian Center for Astrophysics. He chairs the Board on Physics and Astronomy of the National Academies and the advisory board for the Breakthrough Starshot project.

Credit: Nick Higgins